\documentclass{elsart}

\usepackage{graphicx}

\usepackage{amssymb}

\begin{document}

\begin{frontmatter}

\title{Structural effects of ice grain surfaces on the hydrogenation of CO at low temperatures}
\author[label1]{H. Hidaka\corauthref{cor1}}\ead{hidaka@lowtem.hokudai.ac.jp}, \author[label1]{N. Miyauchi}, \author[label1]{A. Kouchi}, \author[label1]{N. Watanabe}
\corauth[cor1]{Corresponding author address.  Institute of Low Temperature Science, Hokkaido University, N19-W8, Kita-ku, Sapporo 060-0819, Japan. FAX: +81-11-706-7142}
\address[label1]{Institute of Low Temperature Science, Hokkaido University, N19-W8, Kita-ku, Sapporo 060-0819, Japan}

\begin{abstract}
Experiments on the hydrogenation of CO on crystalline and amorphous ice at 15~K were carried out to investigate the structural effects of the ice surface. 
The effective rate of H atom addition to CO on the amorphous ice was found to be larger than that on the crystalline ice, while CO depletion on crystalline ice became larger 
after long exposure. 
We demonstrated that the CO-coverage on the ice surfaces dominates the effective reaction rate rather than the surface structure. 
The larger depletion of CO on crystalline ice, as compared to amorphous ice, suggests easier desorption of CO and/or products by the heat of the reaction.
\end{abstract}

\begin{keyword}
\sep amorphous solid water, surface reaction
\PACS 34.90.+q/98.58.Bz
\end{keyword}
\end{frontmatter}

\section{Introduction}
Chemical reactions on solid surfaces are of great interest in the fields of catalytic science, atmospheric chemistry, and astrochemistry. 
In catalytic science, a number of studies on metal surfaces (e.g. Pt, Ag, Au, Cu, Ni, Rh, Mo) have been carried out to investigate the catalytic reactivity. 
Recently, it was reported that the reactivity of surface reactions is affected by the surface morphology not only in metals~\cite{Balint01,Balint02} but also in metal oxides~\cite{Martra01}, 
and semiconductor surfaces~\cite{Ramalingam01}. 
In atmospheric chemistry and astrochemistry, physical and chemical reactions on or in solid water are very important~\cite{Zondlo00,Millar93}. 
Thus, studies have been carried out in many different ways such as photolysis~\cite{dfHendecourt86,Yabushita07}, ion bombardment~\cite{Moor00,Kondo04}, 
electron-stimulated reactions~\cite{Sieger98,Zheng06} and atomic reactions~\cite{Hornekaer03,Perets05}. 
Recently, the influence of the porosity on the production of H$_2$ and O$_2$ in amorphous ice were investigated by the electron-stimulated reaction~\cite{Grieves05}.
However, the influence of the structure of ice (morphology) on the reactivity of chemical reactions on the surface has been still poorly investigated.

The hydrogenation of CO on low temperature solid H$_2$O has been frequently discussed for the formation route of formaldehyde (H$_2$CO) and methanol (CH$_3$OH) 
in astrochemistry, because large amounts of solid H$_2$CO and CH$_3$OH have been observed in the amorphous solid water (ASW) mantle of interstellar dust grains~\cite{Tielens82,Hasegawa92,Charnley97}. 
In our previous experiments, the successive hydrogenation of CO (CO $\rightarrow$ H$_2$CO $\rightarrow$ CH$_3$OH) on the ASW surface was investigated extensively. 
It was demonstrated that reactions strongly depend not only on the temperature~\cite{Watanabe02,Watanabe03} but also on the ice composition~\cite{Watanabe04,Watanabe06}. 
In the present study, the structural effect of the solid water surface on the hydrogenation of CO at a low temperature was investigated. 
The reactivity of CO hydrogenation was compared for ASW and crystalline ice (CI) surfaces.

\section{Experimental}
Experiments were conducted in an ultrahigh vacuum apparatus with a base pressure below 2 $\times$ 10$^{-10}$~Torr, named the Laboratory Setup for Surface reactions in Interstellar Environment (LASSIE). 
LASSIE is equipped with an atomic source, a closed--cycle He refrigerator, a quadrupole mass spectrometer, and a Fourier transform infrared spectrometer (FTIR). 
Details of this apparatus have been described previously~\cite{Watanabe03,Hidaka04}. 
Experiments on the hydrogenation of CO and temperature programmed desorption (TPD) of CO were performed for comparison of the reaction rate constants of hydrogenation 
and the surface area on ASW and CI, respectively. 

Two kinds of layered ices, CO on ASW and CO on CI, were prepared on a cold aluminum substrate by vapor deposition through a capillary plate with a deposition angle of 41$^\circ$. 
The temperature of the substrate was monitored by a Si diode with an accuracy of $\pm$~0.2~K. 
ASW can be formed using a slow rate of vapor deposition on a cold substrate which is below about 100~K, and CI can be also formed at about 120--220~K~\cite{Jenniskens98}. 
In the present experiment, the underlying H$_2$O layer was grown at 15~K and 145~K for ASW and CI, respectively. 
The top CO layer was deposited on the ASW and CI at 15~K. 
The ice samples were exposed at 15~K to a cold atomic hydrogen beam ($\sim$ 100~K) with a flux of 2 $\times$ 10$^{14}$~cm$^{-2}$s$^{-1}$. 
The variation of the surface composition was measured by FTIR after exposure to hydrogen atoms. 
The amount of the underlying H$_2$O in units of column density cm$^{-2}$ was approximately 17 $\times$ 10$^{15}$ cm$^{-2}$ and that of the top CO layer was varied from 0.2 $\times$ 10$^{15}$ to 2 $\times$ 10$^{15}$~cm$^{-2}$. 
Assuming the surface number density of 10$^{15}$ cm$^{-2}$, those column densities correspond to 17, 0.2, and 2 monolayers. 
However, since the ice surface is not perfectly flat, 1 monolayer is not equivalent to unity of coverage. 
Use of term monolayer may mislead the readers. 
Therefore, we use column density instead of monolayers in this article. 
It is well known that the surface area of ASW is larger than that of the CI surface~\cite{Kimmel01}. 
Therefore, CO-coverage of the ASW surface must be lower than that of the CI surface even if the amount of CO deposition is the same. 
For the calculation of the column density, the integrated absorption coefficients were adopted to be 1.1 $\times$ 10$^{-17}$~cm~molecules$^{-1}$ for CO~\cite{Gerakines95}, 
2.0 $\times$ 10$^{-16}$~cm~molecule$^{-1}$ for ASW and 2.1 $\times$ 10$^{-16}$~cm~molecule$^{-1}$ for CI~\cite{Hagen81}.

In order to estimate the ratio of the surface area between ASW and CI, the surface areas were determined by measurements of the monolayer desorption peaks in the TPD spectra. 
In the TPD, the temperature of the substrate was controlled at a rate of temperature rise of 2 $\pm$ 0.5~K~min$^{-1}$ using a ceramic heater mounted on the cold head of the He refrigerator. 
The desorption temperature of CO from the aluminum substrate is close to that from ASW and CI. 
Therefore, if a significant amount of CO sticks on the aluminum substrate, the ice surface areas cannot be measured correctly. 
Thus, the sample ices were prepared by background deposition in which the chamber was filled uniformly with water vapor, in order to prevent CO from adsorbing on the aluminum substrate by the covering of H$_2$O perfectly. 
However, the difference in the deposition method affects the ratio of the surface area, because the morphology (surface area) of ASW strongly depends on the deposition (incident) angle of the H$_2$O molecules 
with respect to the substrate.
In contrast, the morphology of the CI is almost independent of the deposition method~\cite{Kimmel01} due to the surface diffusion of H$_2$O at the higher surface temperature prior to incorporation into CI. 
Accordingly, the measured surface area of ASW grown by background deposition in the TPD experiment should be corrected to that grown by the capillary deposition with the present angle of 41$^\circ$. 
The correction factor $\alpha$ is obtained from the relation between surface areas of amorphous ice formed by the well-collimated beam deposition at many angles and by the background deposition in ref.~\cite{Kimmel01}.
The ratio of surface areas between ASW and CI samples for the hydrogenation experiment can be estimated by the measured surface ratio 
by the CO TPD experiments using the following equation with the correction factor $\alpha$:
\begin{equation}
\frac{S_{\rm ASW}}{S_{\rm CI}} = \alpha \frac{I_{\rm ASW}}{I_{\rm CI}},
\label{f1}
\end{equation}
where $S_{\rm ASW, CI}$ is the area of ASW and CI by the capillary deposition, respectively, $I$ indicates the integrated desorption signals from the first layer of the ice surface in the CO TPD spectrum
from ice samples of the background deposion growth, and $\alpha$ is a correction factor which was obtained to be 0.39.

\section{Result and Discussion}
The top panels in Fig.~1 show the IR adsorption spectra of the initial CO on ASW and CI at 15~K, respectively. 
Many small spikes observed at around 2250--2800 cm$^{-1}$ are attributed to noise due to vibration of the He refrigerator. 
The amount of CO is about 0.4 $\times$ 10$^{15}$~cm$^{-2}$ for both ice samples. 
The broad or split peaks for the OH--stretching mode of H$_2$O at around 3200--3500~cm$^{-1}$ clearly indicate the features of the ASW and CI, respectively~\cite{Hagen81}. 
The bottom panels in Fig.~1 show the variation in the absorption spectra of these ices during exposure to hydrogen atoms. 
The formation of H$_2$CO and CH$_3$OH with decreasing CO were observed in both samples. 
No peaks were observed around the wavenumbers reported for the peaks of HCO, CH$_2$OH in N$_2$ and Ar matrix at 14~K~\cite{Jacox73,Jacox81} and CH$_3$O on low temperature 
surface~\cite{Sim95}. Although the absorption coefficients of those intermediate radicals are unknown, those coefficients are expected to be similar to those for H$_2$CO and CH$_3$OH.
Therefore, we conclude that the signals for those intermediate radicals are under the detection limits. This means that the the reaction rates of H$_2$CO and CH$_3$OH formation 
are much faster than those of HCO and CH$_2$OH or CH$_3$O formation, respectively, because H$_2$CO and CH$_3$OH are formed by very fast reaction, namely radical-radical reactions.
Fig.~2 shows the variations in the column densities of CO, H$_2$CO, and CH$_3$OH under the same experimental condition as in Fig.~1. 

\begin{figure}
\begin{center}
\includegraphics[width=12cm]{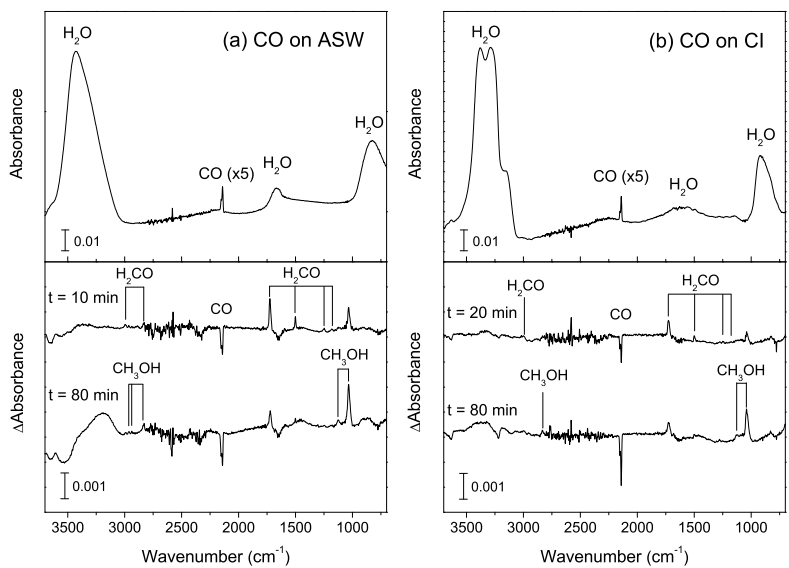}
\end{center}
\caption{\label{fig1} Initial IR absorption spectra ($top$) and variations in the absorption spectra ($bottom$) with H atom exposure for (a) CO on ASW and (b) CO on CI, respectively. 
Peaks below and above the base line represent a decrease and increase in the absorbance, respectively. 
There is vibrational noise at around 2250--2800 cm$^{-1}$ due to the He refrigerator.}
\end{figure}

The column densities were calculated from the integrated absorbances and absorption coefficients of 1.1 $\times$ 10$^{-17}$~cm~molecule$^{-1}$ at 2142~cm$^{-1}$ for CO~\cite{Gerakines95}, 
9.6 $\times$ 10$^{-18}$~cm~molecule$^{-1}$ at 1722~cm$^{-1}$ for H$_2$CO~\cite{Schutte93}, and 1.6 $\times$ 10$^{-17}$~cm~molecule$^{-1}$ at 1032~cm$^{-1}$ for CH$_3$OH~\cite{Kerkhof99}. 
These column densities were normalized to that of the initial CO. 
In both samples, the CO depletion and the production curves of H$_2$CO and CH$_3$OH imply that conversions of CO to H$_2$CO and CH$_3$OH proceed by the successive hydrogenation reaction: 
CO $\rightarrow$ HCO $\rightarrow$ H$_2$CO $\rightarrow$ CH$_3$O (CH$_2$OH) $\rightarrow$ CH$_3$OH on the sample surface. 

The mechanisms of these successive hydrogenation reactions on ASW have been studied in detail previously by our group~\cite{Watanabe02,Watanabe03,Watanabe04,Watanabe06}. 
In the present experimental condition, the irradiated H atoms will react with CO after adsorption or traveling on the sample surface, which is the so-called Langmuir-Hinshelwood or hot atom process~\cite{Hidaka07}.
The depletion curves of CO obviously indicate that reaction rate of CO + H $\rightarrow$ HCO on ASW is faster than that on CI. 
The reaction rate constants for the hydrogenation of CO were obtained from data fitting of the variation of CO by the following equation: 
\begin{equation}
\frac{{\rm [\Delta CO]}_t}{{\rm [CO]}_0} = - A \{ {\rm exp}(-knt)-1 \}, 
\label{f2}
\end{equation}
where $A$ is a saturation value, $t$ is the exposure time, and $n$ and $k$ are the number density of H atoms on the surface and the reaction rate constant, respectively. 
$n$ is assumed to be time-independent and is governed mainly by the balance between the supply of adsorbed atoms from the beam and the loss by the H-H recombination and desorption of H atoms from the surface. 
We could not obtain the value of $k$, because $k$ cannot be separated from the fitting parameter of $kn$ due to the difficulty in estimating the absolute value of $n$. 
Therefore, the reactivity of the hydrogenation was evaluated by using the effective rate constant $k_{\rm eff}$ which is defined as $kn$ in the present study.

\begin{figure}
\begin{center}
\includegraphics[width=6cm]{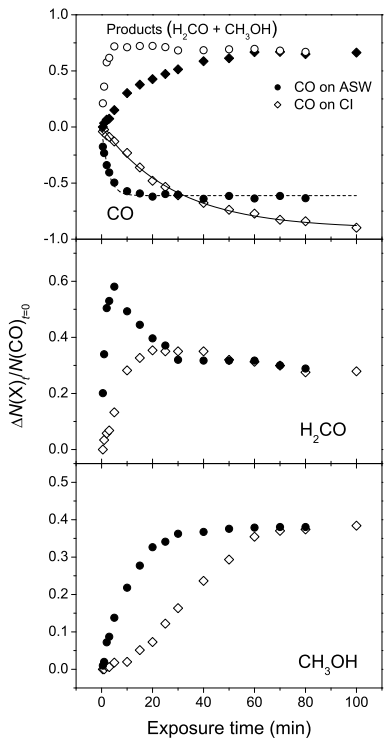}
\end{center}
\caption{\label{fig2} The variations in column densities of CO ($top$), H$_2$CO ($middle$), and CH$_3$OH ($bottom$) as a function of exposure time. 
The experimental condition is same as that in Fig.~1. 
Open circles and solid diamonds in the top panel show the column densities of the reaction products (sum of H$_2$CO and CH$_3$OH) on ASW and CI, respectively. 
The solid and dashed lines represent least-squares fitting of the depletion plots of CO to the single exponential function indicated as Eq. (2).}
\end{figure}

The experiments for the hydrogenation of CO were performed for various initial amounts of CO on CI and ASW (spectra not shown). 
Fig.~3 represents the dependence of the effective reaction rate constants on the initial amount of CO. 
The effective rate constants were determined by the fitting of the CO depletion plots to Eq.~(\ref{f2}) as shown in Fig.~2. 
In both samples, the effective rate constants monotonically decreased with increasing initial amount of CO. 
The rate constants for ASW were always larger than those for CI for any amount of CO. 
Meanwhile, the rate constant for ASW was almost equivalent to that on CI when the initial amount of CO was about 1.84 $\times$ 10$^{15}$~cm$^{-2}$ for ASW and 0.2 $\times$ 10$^{15}$~cm$^{-2}$ for CI. 

There are two possible explanations for the higher reactivity on ASW compared to that on CI, the larger number density of H atoms $n$ on ASW or the larger reaction rate constant $k$ on ASW. 
That is, it is important to clarify which parameters of $k$ and $n$ dominate the higher reactivity on ASW, because the effective rate constant is represented as $kn$. 
The estimation of the value of $k$ is not easy in the present experiment. 
Therefore, we concentrate on the evaluation of $n$. 
The number density $n$ is determined by the balance between the gain by the adsorption of incoming H atoms and the loss by the H-H recombination reaction to form H$_2$ and H atom desorption. 
The rate of H atom desorption can be written as $\nu_0$exp$ \{ -E_{\rm d}/kT \}$, where $E_{\rm d}$ is the adsorption energy and $\nu_0$ is the frequency of vibration, 
indicating that decreasing the adsorption energy sufficiently reduces the number density of H atoms on the surface and finally leads to a lower effective reaction rate. 
In other words, if the desorption rate is significantly high, H atoms desorb before the hydrogenation reaction occurs. 
The adsorption energy of H atoms to ice surface was reported to be about 350--650~K~\cite{Buch91,Al-Halabi07}, while that from solid CO has not been reported. 
However, several experiments have strongly suggested a lower adsorption energy for solid CO~\cite{Watanabe04,Hidaka04}. 
In addition, it is plausible that the surface composed of molecules having higher dipole moment produces the higher binding (physisorption) energy with H atom. 
The dipole moment of H$_2$O is approximately 20 times larger than that of CO.
It is reasonable to consider that CO-coverage of H$_2$O ice dominates the average number density $n$. 
Another possible factor for enhancing the number density is the morphology of the ice. 
ASW has a very irregular surface structure with many cracks~\cite{Dohnalek03} and can trap a large number of molecules in the cracks (pores)~\cite{Ayotte01}. 
Therefore, ASW may have a higher $n$. 
To compare the effects of CO-coverage and the morphology, we measured the ratio of surface areas between ASW and CI and the effective reaction rate with various coverages.

\begin{figure}
\begin{center}
\includegraphics[width=8cm]{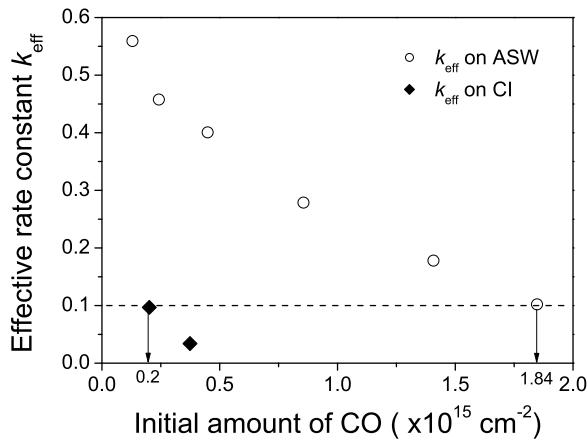}
\end{center}
\caption{\label{fig3} Plots for the effective rate constant $k_{\rm eff}$ for CO + H $\rightarrow$ HCO on ASW (open symbols) and CI (solid symbols). 
$k_{\rm eff}$ is defined by multiplying the reaction rate constant $k$ by the number density of adsorbed hydrogen atom $n$.
The dashed line gives the value of $k_{\rm eff}$ when $k_{\rm eff}$ values are almost equivalent between the ASW and CI surfaces.}
\end{figure}

Fig.~4 shows the TPD spectra of CO at various deposition temperatures in the range of 26--28~K for (a) ASW and 22--28~K for (b) CI, respectively. 
Two desorption peaks were observed in the TPD spectra for ASW and CI at low deposition temperatures. 
The desorption temperature reflects the binding energy between CO and adsorbed sites. 
Since the binding energy of CO-H$_2$O is larger than that of CO-CO, the high and low temperature peaks indicate monolayer and multilayer desorption, respectively, in both ice samples. 
Below the deposition temperature of 27~K, peak areas of multilayer desorption increase with decreasing deposition temperature of CO, whereas those of monolayer desorption were unchanged in the range of the present deposition temperatures. 
This means that the amount of CO that can adsorb on the ice surface below 27~K is proportional to the surface area of the solid H$_2$O. 
Consequently, the ratio of the integrated intensities of the TPD peaks for the monolayer between ASW and CI represents the ratio of surface areas, $I_{\rm ASW}/I_{\rm CI}$. 
From the CO TPD spectra, $I_{\rm ASW}/I_{\rm CI}$ is about 20. 
Finally, using Eq.~(\ref{f1}), the ratio of the surface areas, $S_{\rm ASW}/S_{\rm CI}$, is approximately 8.

The estimated ratio of the surface area indicates that ASW requires 8 times larger amount of initial CO to produce the same coverage of CI. 
If the coverage of CO on ASW and CI were equivalent, the number densities of adsorbed hydrogen atoms should also be equivalent on ASW and CI due to the same ratio of the number of CO sites to that of H$_2$O sites per unit surface area. 
In Fig.~3, the values of $k_{\rm eff}$ on ASW and CI show almost the same values when the amount of initial CO on ASW is about 9 times larger than that on CI. 
This means that the reaction rate constant $k$ is almost independent of the ice structure. 
In other words, the higher reactivity on ASW compared to that on CI for the same initial amount of CO is attributed to a higher density of hydrogen atoms on ASW due to lower CO-coverage compared to CI. 
Recently, it was reported that the binding energy of the trapped H atoms calculated for ASW is 250~K higher than that for CI~\cite{Al-Halabi07}. 
This indicates that the desorption rate of H adatoms for ASW is orders of magnitude lower than that for CI at the same ice temperature. 
We do not exclude this effect on the number density of H adatom on ASW and CI. 
However, our experimental results show that the reactivity of the CO hydrogenation is dominated by the CO-coverage rather than difference in binding energy.

\begin{figure}
\begin{center}
\includegraphics[width=8cm]{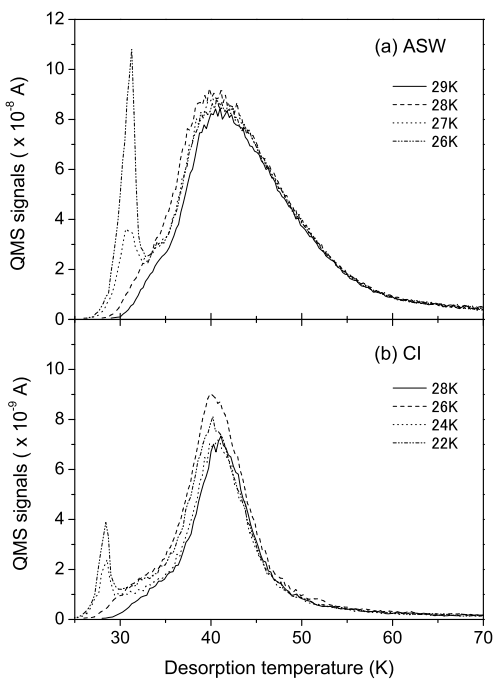}
\end{center}
\caption{\label{fig4} TPD spectra for increasing adsorption temperature of CO from 26~K to 28~K for (a) ASW and 22~K to 28~K for (b) CI, respectively. 
Heating rate of the samples was 2~K~min$^{-1}$. The amount of H$_2$O in ASW and CI was 17 $\times$ 10$^{15}$~cm$^{-2}$. 
Note that the scale of vertical axis of (a) is one order of magnitude larger than that of (b).}
\end{figure}

Finally, the amount of reaction products (open symbol in the top panel in Fig.~2) on CI differs from the amount of depleted CO, although the sum rule between the products and depleted CO is accepted on ASW. 
This discrepancy suggests easy desorption of CO and/or products by the heat of the reaction on CI. 
In the ASW, the desorbed CO and/or products would be re-trapped on the surface by multiple collisions before being released to vacuum because of the irregular (porous) surface structure~\cite{Perets05}.

\section{Summary}
Experiments on the hydrogenation of CO on ASW and CI at 15~K were performed to investigate the structural effects of the ice surface. 
The effective rate constant of the reaction H + CO $\rightarrow$ HCO on ASW was larger than that on CI for the same initial amount of CO. 
The higher reactivity on ASW is attributed to a higher density of adsorbed hydrogen atoms on ASW due to lower CO-coverage compared to CI. 
The amount of reaction products was almost equivalent to the amount of depleted CO on ASW, whereas the amount of reaction products was significant smaller than that of the depleted CO on CI. 
This discrepancy implies that in the case of CI, CO and/or products immediately desorb by the heat of the reaction, while for ASW, the desorbed species may be re-trapped by the irregular surface.

\section{Acknowledgements}
This work was supported in part by a Grant-in-Aid for Scientific Research from the Japan Society for the Promotion of Science and the Ministry of Education, Science, Sports 
and Culture of Japan.

\end{document}